\documentclass[12pt]{article}
\usepackage{setspace}
\usepackage{graphicx}
\usepackage{float}
\usepackage{amsmath}
\usepackage{amssymb}
\usepackage{textcomp}
\usepackage{verbatim}
\usepackage{url}
\title{Refractive index evaluation of porous silicon using Bragg reflectors}

\author{D. Estrada-Wiese$^1$, J.A. del R\'io$^2$\\
$^1$ Instituto de Investigaci\'on en Ciencias B\'asicas y Aplicadas,\\
 Universidad Aut\'onoma del Estado de Morelos, Av. Universidad\\
 No. 1001 Col. Chamilpa, 62209, Cuernavaca Morelos M\'exico.\\
email: de.e.wiese@gmail.com\\
$^2$ Instituto de Energ\'ias Renovables, \\
Universidad Nacional Aut\'onoma de M\'exico,\\
Privada Xochicalco S/N, 62580 Temixco Morelos,
M\'exico\\
email:arp@ier.unam.mx\\ }

\date{}

\begin{document}

\maketitle

\begin{abstract}
There are two main physical properties needed to fabricate 1D photonic structures and form perfect photonic bandgaps: the quality of the thickness periodicity and the refractive index of their components. Porous silicon (PS) is a nano-structured material widely used to prepare 1D photonic crystals due to the ease of tuning its porosity and its refractive index by changing the fabrication conditions. Since the morphology of PS changes with porosity, the determination of PS's refractive index is no easy task. 
To find the optical properties of PS we can use different effective medium approximations (EMA). In this work we propose a method to evaluate the performance of the refractive index of PS layers to build photonic Bragg reflectors. Through a quality factor we measure the agreement between theory and experiment and therein propose a simple procedure to determine the usability of the refractive indices. We test the obtained refractive indices in more complicated structures, such as a broadband Vis-NIR mirror, and by means of a Merit function we find a good agreement between theory and experiment.  
With this study we have proposed quantitative parameters to evaluate the refractive index for PS Bragg reflectors.
This procedure could have an impact on the design and fabrication of 1D photonic structures for different applications.

\end{abstract}

\thispagestyle{empty}

\section*{ Keywords} Refractive Index, Porous Silicon, Effective medium approximation, Bragg reflectors, Photonic structures.\\

PACS: 78.20.-e; 68.65.Ac.

\section{Introduction}
Porous materials are distinguished because their optical characteristics depend strongly on their structural properties (porosity, pore size and pore distribution). Particularly, in materials with air-filled pores, such as porous silicon (PS), the refractive index is directly related to its porosity; however its determination is challenging due to the great variety of microstructures emerging from the diversity of the fabrication specifications.

PS is usually fabricated by electrochemical etching of crystalline Si in a hydrofluoric acid solution, thus the resulting nanostructure is composed of Si and air in a sponge-like structure. This process regulates the generated porosity by changing the applied current density and then modifying the refractive index of the resulting PS. This fact opens the possibility of fabricating 1D photonic structures with PS, where controlling the propagation of light in a dielectric medium is sought. Bragg mirrors, microcavities, filters,  bio and chemical sensors  are some of the simplest photonic multilayer arrangements that can be fabricated  with PS \cite{filters,reviewPS}. For example, a Bragg-reflector is composed of a periodic stack of layers which alternate between high and low refractive indices with a high contrast between layers. If each layer satisfies the quarter wavelength condition, a selective mirror that reflects a central wavelength can be constructed \cite{HSohn,oxiBragg,distributed-Bragg-reflector,tailoring-ARP}. 

In order to build efficient and high quality 1D photonic structures the refractive index of PS must be determined accurately. For instance, PS biosensing devices are based on the change of the effective refractive index due to the presence of molecules in the nanostructure modifying its spectral response \cite{structural,biosensor}, thus determining accurate refractive index values of the porous material is essential.

Most of the current research has been done using constant refractive indices \cite{Optik-SEM,drug-delivery,Negative,low-N-contrast} or, when not considering constant values an arbitrary dependency that adjusts the experimental behaviour are often proposed \cite{oxiBragg,OpticalSensors,Optik-Ajustes}. There are many examples presenting slight differences between theoretical and experimental results that need deeper explanations \cite{Optik-SEM,Cisneros-elips,KK}. This matter can be surpassed by measuring the effective refractive index as a function of the wavelengths, for example by spectroscopic ellipsometry (SE). Although there are recent reports where SE measurements are used to determine the refractive index of PS \cite{Zeuz,elipsometria,petrik,otroSE}, in this work we found that this technique is not adequate for our purposes (see Supplementary Appendix).

Hence, in this study we calculated the PS refractive indices using effective medium approximation (EMA) methods and evaluated the usability of these data by using them to predict the reflectance of fabricated Bragg mirrors. We obtained the theoretical reflectance spectra using the transfer matrix method and compared them to the experimental measurements. This comparison provides a good assessment for the refractive index determination where several proposals were evaluated by defining a $Q_e$-factor in the analysis of the Bragg reflectance spectra. In this manner we offer a quantitative strategy to select adequate effective refractive indices for the construction of high reflective and broadband Bragg mirrors.

In what follows we present the fabrication procedure of PS and its porosity characterization. With these values we approach different refractive indices using several EMAs. Subsequently, we evaluate the performance of the refractive indices by fabricating Bragg reflectors centered at different wavelengths and calculating the $Q_e$-factor for each one. Afterwards we examine the use of the most adequate refractive indices and fabricate a broadband mirror to ultimately test their usability. Finally we present the concluding remarks.

\section{Experimental details}\label{ps}

In this section we present the fabrication procedure of the PS samples and detail the method we used to characterize its porosity and thickness. These values were later used for the determination of the refractive index of PS. 
The PS samples for this study were fabricated by an anodic electrochemical dissolution of highly boron-doped p$^{+}$-type (100) crystalline silicon (c-Si) wafers with resistivity $<$ 0.005 $\Omega \cdot cm$. In order to ensure electrical conduction during anodization, an aluminum film was deposited on the backside of the c-Si substrates and then heated at 500 $^{\circ}$ C during 30 min in nitrogen atmosphere. The substrates were electrochemicaly etched in an electrolyte composed of ethanol, HF and glycerin in a volume ratio of 7:3:1 (if the total volume of the electrolyte is 55 ml then we use 35 ml of ethanol, 15ml of HF and 5 ml of glicerol). We fabricated single high porosity layers by applying a current density of 40.0 $mA/cm^2$ and low porosity layers by using 3.0 $mA/cm^2$. After electrochemical etching, the samples were rinsed in ethanol for 10 minutes and dried under a nitrogen stream. We subsequently oxidized the samples for stabilization of the PS at 300$^{\circ}$ C during 15 minutes.

To measure the porosity produced with these current densities we fabricated $5000$ nm thick films and used the gravimetric method \cite{grav} where the silicon wafer to be etched is weighted before anodization ($m_{1}$), immediately after anodization ($m_2$), and after dissolving the PS layer in an aqueous solution of sodium hydroxide ($m_3$) using the formula: 
\begin{equation}
\displaystyle P= \frac{(m_1 - m_2)}{(m_1-m_3)}.
\end{equation}

\begin{figure}[ht]
\centering
\includegraphics[width=1\linewidth]{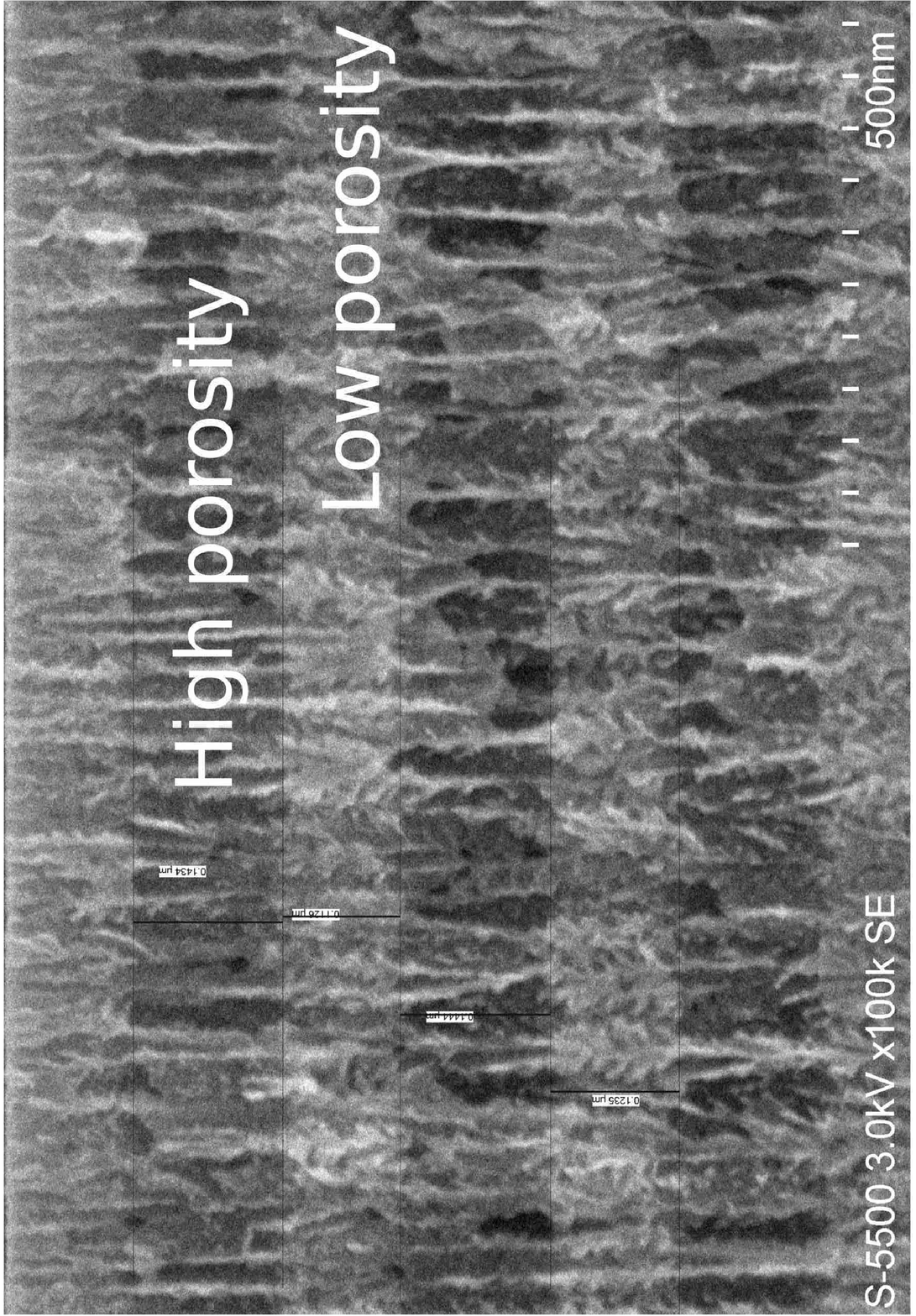}
  \caption{ SEM image. SEM image showing the two different porosity layers $P_a$ and $P_b$, used in this work. } 
\label{SEM}
      \end{figure}

In this manner, measuring the corresponding samples with a Sartorius Microbalance (model MC 5) with a precision of 0.0005 mg,  we calculated porosities of $P_a= 79.2 \% $  and $P_b= 59.4 \% $ (from standard error propagation the error percentage is less than $0.15 \%$ for each different porosity sample), where the subscript $a$ and $b$ stand for high and low porosity layers. In addition, we characterized the thickness of the formed PS films using cross sectional SEM images using a Hitachi S5500 electron microscope (see figure \ref{SEM}) and determined the etch rate as $v_a= 14.49$ $nm/s$ and $v_b= 1.72$ $nm/s$. With these values we calculated the time at which each current density needed to be applied to form the desired thickness of the layers. It is a well known issue that the HF concentration decreases with time and layer depth during PS fabrication. To overcome this problem, we implemented 1 second long pauses to the etching time so that the HF concentration can restore and minimize the porosity gradient.

Since the porosity of PS determines its the refractive index and we calculated the porosity values of each PS layer, then we were able to predict its average refractive index using EMA methods. In the next section we present the main effective methods we used for approaching the refractive index of PS.

\section{Effective medium approaches for PS} \label{metodos}

Traditionally the envelope method or the Fresnel's equation are used to obtain the refractive index from the measured reflectance and transmittance spectra of PS monolayers. However, these methodologies present restrictions when the material has high optical absorption or scattering effects, such as PS in the visible range \cite{Arenas}.
 
In particular, the structure and morphology of PS changes as a function of porosity and because the size of its pores is much smaller than the light wavelength it can be described as an effective medium. The high and low porosity layers presented in figure \ref{SEM} show a coral-like structure in the $P_a$ layers and an interlaced branched formation of Si in the $P_b$ layers. These morphologies are complicated and to describe their effective behaviour different EMAs can be used. Many of these methods have been used for the determination of the refractive index but have been chosen arbitrarily as discussed in \cite{expPilon}. Since we do not have enough information to distinguish the agreement with these EMAs and the interaction between light and PS, in this study we evaluated which method is best suited to approach the refractive indices of PS of high and low porosities. For this purpose we selected different EMAs and provided a quantitative method to determine which EMA has the best performance by fabricating Bragg reflectors.  

The effective medium approaches were developed to obtain theoretical values of the effective dielectric function, whose real part $\epsilon_{r}$ relates to the real effective refractive index as $\epsilon_r \sim n_{eff}^2$ when considering low absorption. The following EMAs have been expressed in terms of $n_{eff}$ \cite{EMAs,Wolf, modelosefectivos}, from which the Maxwell-Garnett \cite{MG}, Looyenga \cite{Looy}, the formula of del R\'io et al. \cite{Zimmerman} and Bruggeman \cite{Brugge} stand out among others like the simple linear \cite{Wolf} or parallel interpolations \cite{EMAs}. The linear approximation is calculated using the porosity $P$ that indicates the volume fraction of air in silicon: 
\begin{equation}
 n_{eff} (\lambda)= Pn_{air}(\lambda) +(1-P)n_{Si}(\lambda) ,
\end{equation}
where $n_{Si}$ is the refractive index of silicon and $n_{air}$ the refractive index of air. Whereas the parallel interpolation is calculated as: 
\begin{equation}
\displaystyle \frac{1}{ n_{eff}(\lambda)}= \frac{P}{n_{air}(\lambda)} + \frac{1-P}{n_{Si}(\lambda)}.
\end{equation}

The Looyenga model is best suited for high porosities and is defined as:
\begin{equation}
 n_{eff}^{2/3}(\lambda) = (1-P)n_{Si}^{2/3}(\lambda) +Pn_{air}^{2/3}(\lambda).
\label{looyenga}
\end{equation}
Because the Maxwell-Garnett formula considers isolated spherical particles, where percolation of PS is not contemplated, this model is not relevant for this material \cite{Wolf,optical}. As an alternative to this methods we used the formula proposed by del R\'io et al. (dRZW) based on the Keller reciprocity theorem for effective conductivity in a composed material \cite{Zimmerman}. dRZW considers no particular inclusion shape, so it could be applied to  materials with arbitrary microstructure. We used this formula for the effective refractive index as:
\begin{equation}
 \displaystyle n_{eff} (\lambda)= n_{Si}(\lambda) \frac{1+P(\sqrt{\frac{n_{air}(\lambda)}{n_{Si}(\lambda)}}-1)}{1+P(\sqrt{\frac{n_{Si}(\lambda)}{n_{air}(\lambda)}}-1)}.
\label{zimmerman}
\end{equation}

The Bruggeman approximation is one of the most used EMA for the refractive index determination of PS \cite{EMAs,brugg}. The symmetric Bruggeman considers different sizes of spherical inclusions embedded in a continuos medium and is applicable to any porosity:
\begin{equation}
 \displaystyle P \frac{n_{air}^2(\lambda)-n_{eff}^2(\lambda)}{n_{air}^2(\lambda)+2n_{eff}^2(\lambda)} + (1-P)\frac{n_{Si}^2(\lambda)-n_{eff}^2(\lambda)}{n_{Si}^2(\lambda)+2n_{eff}^2(\lambda)}=0.
\end{equation}

The complex refractive index is defined as $\eta(\lambda)= n(\lambda)-ik(\lambda)$ where absorption is related to the extinction coefficient $k(\lambda)$. The nonsymmetric Bruggeman approximation can be used for the calculation of $k(\lambda)$ of PS \cite{EMAs}:
\begin{equation}
 \displaystyle \frac{k^2(\lambda)}{k_{Si}^2(\lambda)}-\frac{k_{air}^2(\lambda)}{k_{Si}^2(\lambda)}= (1-P) \Bigg[\bigg(1-\frac{k_{air}^2(\lambda)}{k_{Si}^2(\lambda)}\bigg)\bigg(\frac{k^2(\lambda)}{k_{Si}^2} \bigg)^{\frac{1}{3}}\Bigg]. \label{keff}
\end{equation}
Here we determined the extinction coefficient values for high and low porosity PS layers from equation (\ref{keff}) and calculated the refractive indices, $n_a (\lambda)$ and $n_b (\lambda)$ respectively, using the following EMAs: Looyenga model, dRZW formula, the linear and parallel interpolations and the symmetric Bruggeman model. Values for $n(\lambda)$, $k(\lambda)$ of Si and air were taken from \cite{si,aire}.  Although Si at optical frequencies is dependent on carrier concentration, the change of refractive index varies as $\sim 10^{-3}$, which is negligible compared to the change of the refractive index due to the uncertainty in the porosity measurements \cite{carriers}.

\begin{figure}[ht]
\centering
\includegraphics[width=1\linewidth]{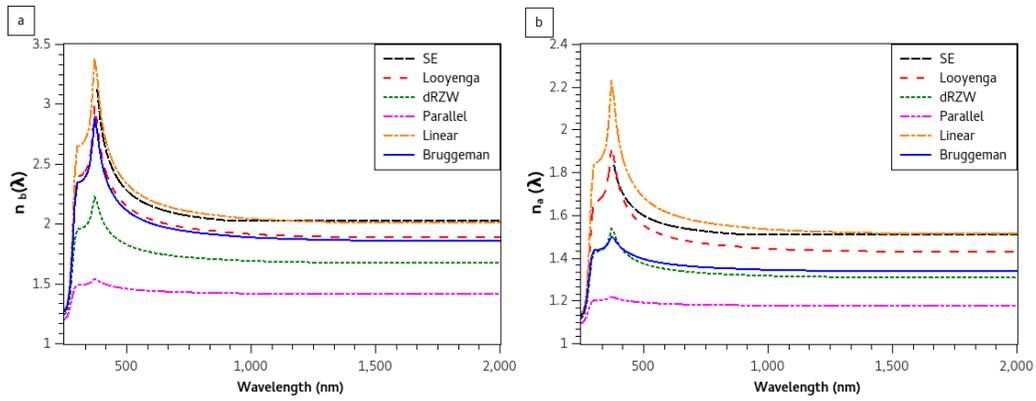}
  \caption{ Refractive index spectra. Refractive index spectra of PS obtained with SE (black line), Looyenga (red line), the dRZW formula (green line), the paralell (pink line), the linear interpolation (orange line) and the Bruggeman approach (blue line), for a) the low porosity layers $P_b$, and b) the high porosity layers $P_a$.} 
\label{comp}
      \end{figure}

In figure \ref{comp} we show a comparison between the spectra of the refractive indices obtained by these effective models and the ones obtained from SE (see Supplementary Appendix for more details). 
We can observe slight differences between the refractive index values obtained from each methodology, therefore we need an efficient procedure to determine which values are the most adequate. For this reason, we fabricated several Bragg mirrors centered at specific wavelengths ($\lambda_0$) using the different formulas for the refractive indices and assess them to evaluate the performance of the refractive indices.

\section{PS Bragg reflectors}\label{bragg}

\begin{figure}[ht]
\centering
\includegraphics[width=1\linewidth]{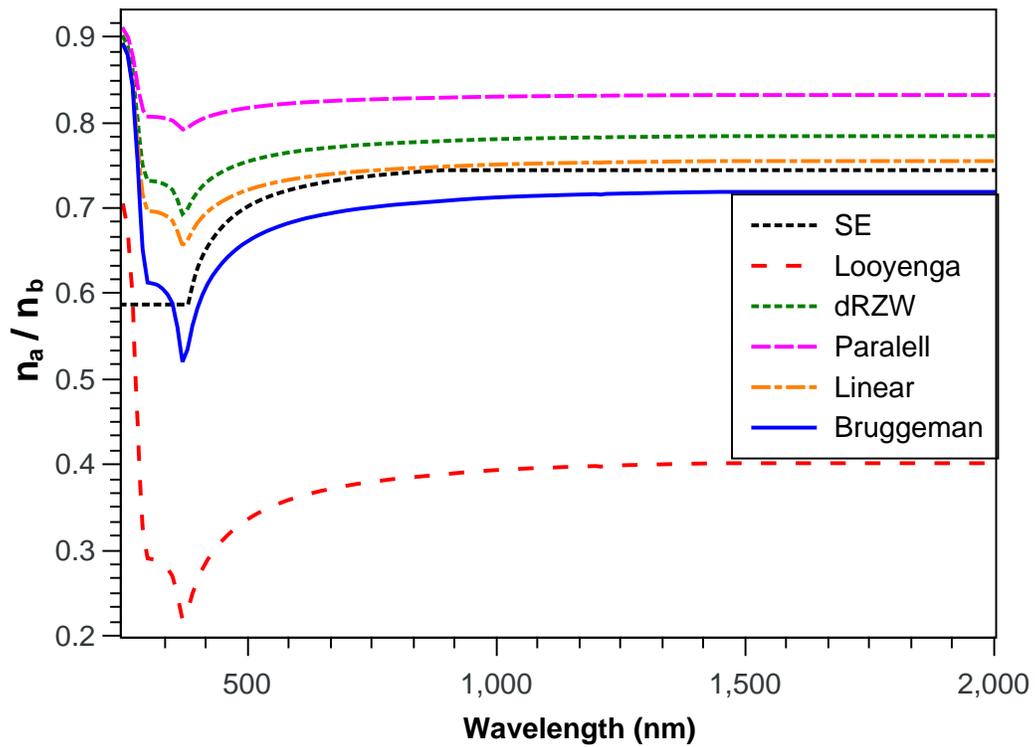}
  \caption{Refractive index rates. Refractive index rates (n$_a$/ n$_b$) obtained from SE measurements (black line), dRZW (red line), the Looyenga model (green line), the paralell (pink line), the linear interpolation (orange line) and the Bruggeman approach (blue line).} 
\label{contraste}
    \end{figure}

Bragg reflectors are the simplest 1D photonic structures, since they are formed of alternating layers of high ($n_a$) and low ($n_b$) refractive index and repeating thickness $d_a$ and $d_b$ respectively. When an electromagnetic wave with a specific wavelength $\lambda_0$ enters the structure, it is partially reflected at each layer interface and satisfies the optical path relation: $n_id_i=\lambda_0/4$, where $i= a$ for the $P_a$ layers and $i=b$ for the $P_b$ layers. Due to the periodicity of the refractive indices, these multiple reflections interfere destructively avoiding the further propagation of the wave. In this manner a forbidden band gap around a central wavelength $\lambda_0$ is formed, i.e. a perfect mirror at $\lambda_0$. The multilayered structure that satisfies these properties is called a Bragg reflector. The procedure to fabricate these mirrors constrains the need of a low index rate ($n_a/n_b$) between the layers in order to have a periodic potential and consequently an increased band gap \cite{tailoring-ARP, index,review1D}. We know from our experimental experience that our $P_a$ and $P_b$ PS fabrication conditions present a high index contrast and by using these we have been able to produce different photonic structures \cite{index,pade,noemi}. Since it is our aim to fabricate high quality 1D photonic mirrors we need to ensure that the refractive indices obtained from the different methods represent effectively the interaction between light and PS. Therefore we calculated all the refractive index rates and present their comparison in figure \ref{contraste}. Here we outstand the Looyenga and the Bruggeman refractive indices which exhibit the lowest rate, indicating that their index contrast is large enough to reproduce the adequate photonic quality we have observed in previous reports.

The thickness and refractive index of the periodic layers that constitute a Bragg mirror determine the $\lambda_0$ it reflects by means of the optical path relation. If for example we design a Bragg mirror centered at $\lambda_0= 600 $ nm we only need to calculate $d_a$ and $d_b$ of the high and low porosity layers by using the values of $n_a(\lambda_0)$ and $n_b(\lambda_0)$ respectively. If we fabricate this mirror and its experimental central wavelength $\lambda_E$, measured from the experimental reflectance spectra, does not correspond to the original $\lambda_0=600$ nm, then we say that the Bragg mirror is shifted to either smaller or larger wavelengths. Moreover this shift means that the values of $n_a(\lambda_0)$ and $n_b(\lambda_0)$ that we used are not the ones that describe accurately the fabricated porosities of PS. 
Thus, with correctly characterized parameters ($d_i$ and $n_i$) we can produce selective Bragg mirrors and use them for several applications. In this work we present a simple  procedure to determine the refractive index of PS based on the performance of their use for fabrication of PS Bragg mirrors, evaluating the concordance between their experimental and theoretical reflectance spectra.

Hence we fabricated several Bragg reflectors in order to validate the performance of the refractive indices that we obtained from each methodology (figure \ref{comp}). Using these values we designed sets of three selective mirrors centered at $\lambda_0 = 600$, $800$ and $1000$ nm respectively, formed of 15 bilayers of high and low refractive indices. The thickness of each layer satisfies the quarter wavelength condition, $d_i= \lambda_0/4n_i$, and is controlled experimentally with the etching time. It is important to notice that each set of Bragg mirrors designed with their corresponding refractive indices (obtained from each method) have different layer thicknesses and because the fabricated porosities ($P_a$ and $P_b$) are the same for all the mirrors, their experimental reflectance spectra will not be equal, even though they are designed to reflect the same $\lambda_0$. We expect that some of them present better agreement with their respective theoretical calculation.

We simulated the theoretical reflectance spectra of the Bragg mirrors using the well known transfer matrix method \cite{Hecht} where we considered the absorption in the multilayers by using the complex refractive index in the calculations. Then we compared the spectra with the  experimental measurements performed with a spectrophotometer UV-Vis-IR (Shimadzu UV1601), see figures \ref{elips},\ref{looyengas} and \ref{effectivos}. The spectrophotometer uses an Aluminum mirror as reference, this is the standard method we have used in our studies.

If the $n_b(\lambda_0)$ and $n_a(\lambda_0)$ used to design and fabricate the mirrors are adequate, then each experimental Bragg reflectance spectrum will correspond to its theoretical counterpart.  We note that in this work we do not consider scattering and absorption effects because they do not affect the central wavelength of the Bragg structure. Therefore, if the spectra are in good agreement, it is possible to validate the usability of the refractive index values using a common metric that determines the quality of the fabricated Bragg structures. In this work we propose the $Q_e$-factor as the theoretical ($\lambda_0$) and experimental($\lambda_E$) central wavelength difference divided by the full width at half maximum ($a_T$) of the theoretical band gap:
\begin{equation}
\displaystyle Q_e= \frac{\mid \lambda_0 -\lambda_E \mid}{a_T}. \label{Q}
\end{equation}
The smaller the $Q_e$-factor, the more accurate the refractive indices which characterize the photonic Bragg structure. We measured the $\lambda_E$ and $a_T$ as the half width at 50$\%$ reflectance. In tables \ref{tabla600}-\ref{tabla1000} we show the $Q_e$-factor calculated for each set of mirrors that we fabricated using the different refractive indices.

\section{Results and discussion}

\begin{figure}[ht]
\centering
\includegraphics[width=1\linewidth]{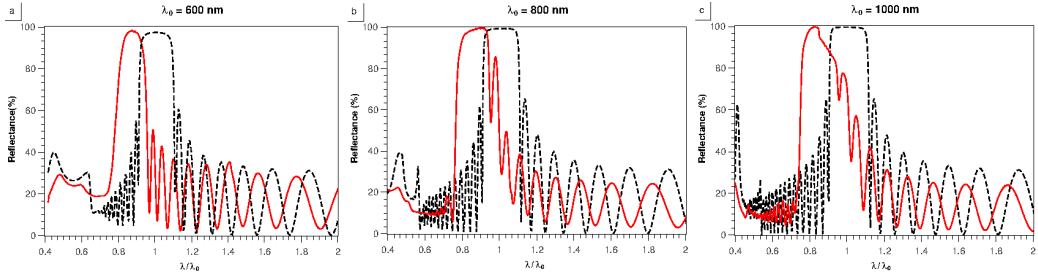}
  \caption{ Reflectance spectra of the SE Bragg mirrors. Theoretical (black dashed line) and experimental (red line) reflectance spectra of PS Bragg mirrors, centered at a) $\lambda_0$= 600 nm, b) 800 nm and c) 1000 nm, using the refractive indices obtained from SE. } 
\label{elips}
    \end{figure}

\begin{figure}[ht]
\centering
\includegraphics[width=1\linewidth]{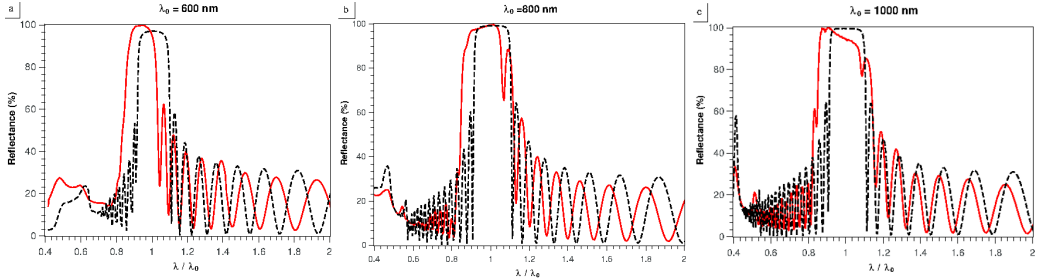}
  \caption{Reflectance spectra of the Looyenga Bragg mirrors. Theoretical (black line) and experimental (red line) reflectance spectra of PS Bragg mirrors, centered at a) $\lambda_0$= 600 nm b) 800 nm and c) 1000 nm, using the refractive indices obtained from the Looyenga method.} 
\label{looyengas}
    \end{figure}

\begin{figure}[ht]
\centering
\includegraphics[width=1\linewidth]{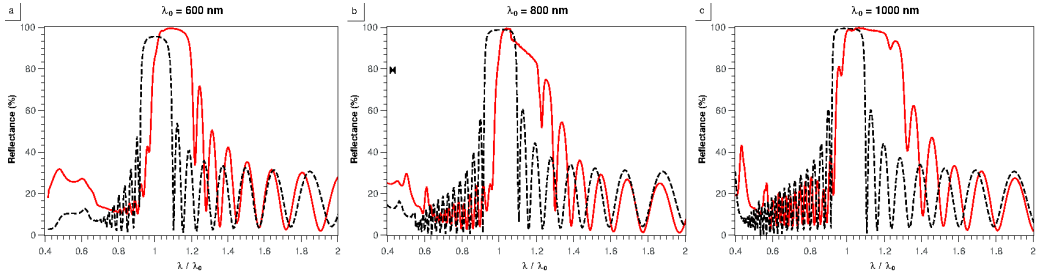}
  \caption{Reflectance spectra of the dRZW Bragg mirrors. Theoretical (black line) and experimental (red line) reflectance spectra of PS Bragg mirrors, centered at a) $\lambda_0$= 600 nm, b) 800 nm and c) 1000 nm, using the refractive indices obtained from dRZW formula.} 
\label{effectivos}
    \end{figure}

Each set of Bragg mirrors produced to validate the performance of the refractive index values are presented here. First, we fabricated three reflectors using the values obtained from SE measurements (for details see Supplementary Appendix) and compared their experimental reflectance spectra to their corresponding theoretical spectra in figure \ref{elips}. Here a shift to shorter wavelengths can be observed in all the experimental spectra which suggests that the SE refractive indices are not adequate. The calculations of the $Q_e$-factor for these mirrors confirm the inadequacy of the SE measurements (see tables \ref{tabla600}-\ref{tabla1000}). The difference between theory and experiment can be understood as a result of the complexity of modelling the PS nanostructure and the use of accurate SE data with complex models to estimate refractive indeces which afterwards are used in simple photonic structures.\\

\begin{figure}[ht]
\centering
\includegraphics[width=1\linewidth]{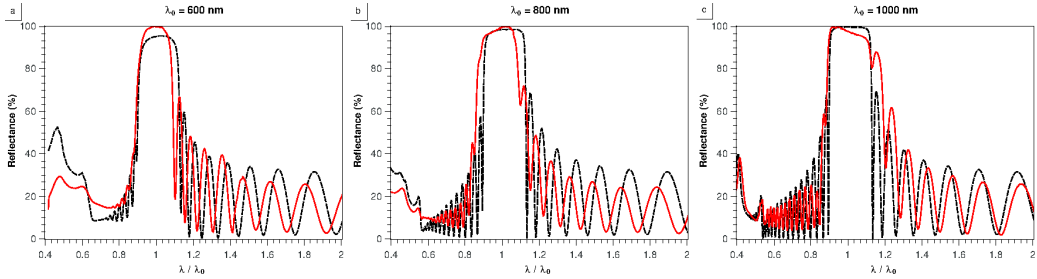}
  \caption{Reflectance spectra of the Bruggeman Bragg mirrors. Theoretical (black line) and experimental (red line) reflectance spectra of PS Bragg mirrors, centered at a) $\lambda_0$= 600 nm, b) 800 nm and c) 1000 nm, using the refractive indices obtained from the Bruggeman approximation.} 
\label{Bruggs}
    \end{figure}

The blue-shift of the experimental reflectance spectra in figure \ref{elips} advises that the refractive index values must be smaller in order to displace the spectra to larger wavelengths to fit the theoretical spectra. Therefore we chose the Looyenga, the dRZW and the Bruggeman refractive indices for the fabrication of other sets of Bragg reflectors following the same mentioned methodology. In figures \ref{looyengas}-\ref{Bruggs} we show the comparison between the theoretical and experimental reflectance of each mirror that were designed using the refractive index values obtained from i) the Looyenga effective model, ii) the dRZW formula and iii) the Bruggeman approximation. 

The best fit between the theoretical and experimental reflectance spectra are the ones fabricated using the Bruggeman refractive indices. Furthermore, we calculated the $Q_e$-factor using equation (\ref{Q}) for these results and present them in tables \ref{tabla600}-\ref{tabla1000}. From these values we can conclude that the Bruggeman refractive indices are the most adequate for the fabrication of the PS Bragg reflectors. Note that the qualitative and quantitative agreement is clearly better that with the other formulas.

\subsection{Broadband mirror} \label{broadband}

\begin{figure}[ht]
\centering
\includegraphics[width=1\linewidth]{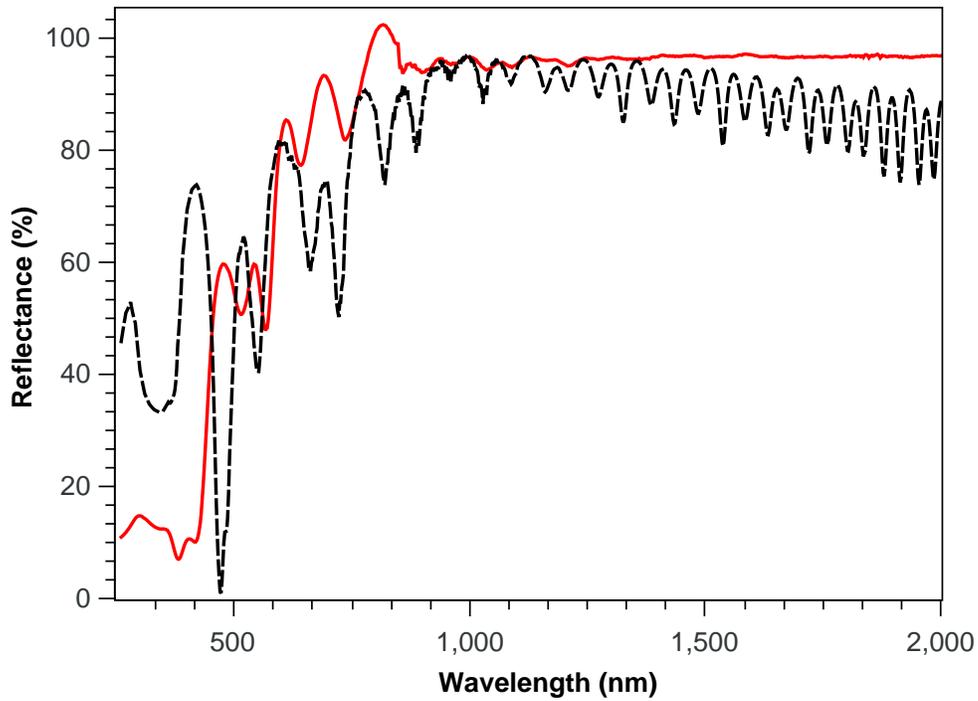}
  \caption{Reflectance spectra of broadband mirror. Theoretical (black line) and experimental (red line) reflectance spectra of a PS broadband mirror using the Bruggeman refractive indices.} 
\label{oldpadeB}
    \end{figure}

The bandwidth of one Bragg reflector represents the wavelength range presenting high reflectance. This range depends on the refractive index rate between high and low porosity layers.  For example, the reflectance spectra of the Bragg mirrors centered at $\lambda_0=600$ nm presented above, show different bandwidths (see $a_T$ values in table \ref{tabla600}). Even though the mirrors are designed to reflect the same $\lambda_0$, they do not have the same bandwidth because the refractive index rates are not the same.
When an enlargement of the photonic band gap is sought, for instance in solar concentration devices \cite{Bety}, the superposition of several Bragg reflectors with a high index rate is needed \cite{pade,OM-mirrors}. From the previously shown results we concluded that the Bruggeman refractive indices were the most adequate to produce Bragg reflectors, and because these also present a low index rate we now use these values to fabricate a broadband mirror of PS. To do so we superpose a selective number of Bragg mirrors (named submirrors) to reflect over a wide wavelength range. However, finding the optimal configuration of submirrors and its $\lambda_0$ to fabricate a high reflecting broadband mirror is not easy. In previous work we reported a simple procedure to determine the $\lambda_0$ of the submirrors based on the  Pad\'e approximant \cite{pade}. Following that methodology we fabricated  a Vis-NIR broadband mirror made of PS and defined the wavelength range from $\lambda_1$ = 400 nm to $\lambda_f$ = 2000 nm. 
We prepared the PS broadband mirror, composed of 20 submirrors of 5 periods, using the Bruggeman refractive indices and the same fabrication conditions as before. The mirror was afterwards oxidized for stabilization at 300$^{\circ}$ C for 15 minutes.
In figure \ref{oldpadeB} we report its theoretical and experimental reflectance spectra and observe a very good fit within the desired wavelength range (400 - 2000 nm). We find that the experimental photonic band gap is widen to larger wavelengths, probably due to the porosity gradient within the multilayered structure. This is a problem that we have encountered in recent studies \cite{pade} and needs further investigation, however we do not address this matter in the present report.

To measure quantitatively the concordance between theory and experiment and since we can not use the $Q_e$- factor criterion for a multi-Bragg structure, we propose a merit function given by:
\begin{equation}
N= \sqrt{\frac{\displaystyle \int_{\lambda_1}^{\lambda_f} {(S(\lambda)-E(\lambda))^2} d\lambda}{(\displaystyle \int_{\lambda_1}^{\lambda_f}{S(\lambda) d\lambda})^2}}, \label{int}
\end{equation}
where $S(\lambda)$ and $E(\lambda)$ are the simulated and experimental reflectance spectra, respectively. Here $N$ must be small for experiments and theory to agree. We calculated the merit function for the broadband mirror and obtained the value $N=0.0047$. This result clearly indicates that the Bruggeman refractive indices are the most adequate for the design and fabrication of PS Bragg and broadband mirrors.

\section{Conclusions}\label{final}

In this study we theoretically and experimentally characterized the refractive index of high (P$_a= 79.2 \%$) and low (P$_b= 59.4\%$) porosity PS films. Through SE measurements and EMA methods, where different fitting models were considered for each porosity, we determined the refractive index of PS.
 
In this work we report a simple procedure to assess the usability of the refractive index values based on the fabrication and evaluation of photonic mirrors. Here we validated the performance of the refractive indices by fabricating Bragg reflectors centered at 600, 800 and 1000 nm and compared the reflectance spectra to theoretical simulations using the $Q_e$- factor criterion. We found that the refractive indices obtained by SE were not adequate and calculated afterwards the refractive indices of PS using EMA, in particular using the Looyenga method, the dRZW formula and the Bruggeman approximation.  Furthermore, we fabricated a broadband Vis-NIR mirror and measured the concordance between theory and experiment using a Merit function. Within this analysis we proved a good agreement for the mirrors produced with the Bruggeman refractive indices and concluded that these are the most adequate values for the fabrication of PS 1D photonic structures presented in this work, such as Bragg or broadband mirrors. With 
this study we have proposed quantitative parameters to evaluate the performance of the refractive index in PS through photonic structures. 
 
\section*{Acknowledgements}

The authors gratefully acknowledge Dr. Z. Montiel-Gonz\'alez for his support on the SE fitting model, M.C. J. G\'omez-Ocampo for helpfull discussions and stable matrix calculations. They also thank J. Campos for SEM images. This work received partial support by Consejo Nacional de Ciencia y Tecnolog\'ia under proyect ``Fronteras de la Ciencia'' 367.

%



\section*{Appedix: Spectroscopic Ellipsometry measurements}
In the appendix we detail the procedure that we used for the determination of the refractive index of the PS high and low porosity samples by Spectroscopic Ellipsometry.

\section*{Figures and Tables}

\begin{table}[ht]
\centering
\begin{tabular}{l l l l}
 
{\bf Methods:} &{\bf $\lambda_E$ (nm)} & {\bf$a_T$ (nm)} & {\bf$Q_e$-factor}  \\ 

SE & 521.8 & 114.7 & 0.6831\\

Looyenga & 593.1 & 110.0 & 0.0624\\

dRZW & 656.6& 104.7 & 0.5404 \\

Bruggeman & 593.6 &137.1 & 0.0466\\ 

\end{tabular}
\caption{  \label{tabla600} \footnotesize The $Q_e$-factor calculated for each Bragg reflector centered at $\lambda_0= 600$ nm fabricated with different refractive indices obtained from SE and EMAs.}
\end{table}

\begin{table}[ht]
\centering
\begin{tabular}{l l l l}
 
{\bf Methods:} &{\bf $\lambda_E$ (nm)} &{\bf $a_T$ (nm)} & {\bf$Q_e$-factor}\\ 

SE & 705.7 & 159.0 & 0.5928 \\ 

Looyenga & 785.2 & 157.0 & 0.0943\\

dRZW & 905.0 & 143.7& 0.7305 \\

Bruggeman & 797.7 & 186.2 & 0.0124\\ 
\end{tabular} 
\caption{\label{tabla800} \footnotesize The $Q_e$-factor calculated for each Bragg reflector centered at $\lambda_0= 800$ nm fabricated with different refractive indices obtained from SE and EMAs.}
\end{table}

\begin{table}[ht]
\centering
\begin{tabular}{l l l l}
 
{\bf Methods:} &{\bf  $\lambda_E$ (nm)} &{\bf $a_T$ (nm)} &{\bf $Q_e$-factor}  \\

SE & 880.3 & 215.2 & 0.5564 \\

Looyenga & 933.1 & 183.4 & 0.3647\\

dRZW & 1127.0 & 180.9& 0.7038 \\

Bruggeman & 1028.1& 231.0 & 0.1216\\ 

\end{tabular} 
\caption{ \label{tabla1000} \footnotesize The $Q_e$-factor calculated for each Bragg reflector centered at $\lambda_0= 1000$ nm fabricated with different refractive indices obtained from SE and EMAs.}
  \end{table}

\section*{Appendix: Spectroscopic Ellipsometry measurements}
\setcounter{figure}{0}
\renewcommand{\thefigure}{A\arabic{figure}}

In this appendix we detail the procedure we used for the determination of the refractive index of high and low porosity PS samples by Spectroscopic Ellipsometry (SE). 

SE is a semiempirical approach based on the measurement of the light polarization transformation that occurs after a polarized beam is reflected from a material. The SE data are adjusted to fit a model that considers the characteristic properties of the material and from which the complex refractive index is calculated \cite{Zeuz,elipsometria}.

In this work we fabricated $500$ nm thick PS monolayers for the SE measurements, which were carried out on a $\alpha$-SE, J.A. Woollam Co spectroscopic ellipsometer. The data analysis was performed using the Complete EASE software where in the fitting model we established the PS as an effective Bruggeman medium (EMA) formed of Si and air. The parameters, such as thickness and porosity, are varied/fitted in the manner that the adjustment between simulated and experimental data present the minimum mean square error (MSE).

As it can be observed in the SEM image (figure 1 of the main text), the morphology of PS is complicated and varies with porosity. The structure of the low porosity layers resembles interlaced branches of Si, whereas the high porosity layers have a coral-like formation which present percolation. These different structures might be represented by different fitting models, such as isotropic, anisotropic or porosity graded. In this work we tried the above mentioned possibilities for the $P_b$ layers and found the minor MSE values using the graded porosity model. Because of the coral-like structure observed in the $P_a$ layers we used an anisotropic model for the fitting of these layers. The optical graded and anisotropic models are commonly used to approach the optical properties of PS when using SE \cite{Zeuz, elipsometria} and in this study we used them as follows:

a) For the low porosity layers we defined a depth dependent porosity gradient by simulating $10$ isotropic sublayers with a gradual porosity. This graded layer is on top of another layer with lower porosity and placed on the Si substrate. The fitting through the software gave a thickness of 493 nm on average, whereas from the SEM measurements we obtained a thickness of 511 nm. The MSE values obtained for this model range between 35 and 40.

b) For higher porosities we approached a coral-like structure where a directional dependency of the refractive index was generated and therefore considered an uniaxial anisotropic layer \cite{Zeuz}. The software gave a fitted thickness of 464 nm on average, which compared to the SEM measurements where we obtained a thickness of 507 nm, the SE fitting presented higher differences. 
The fitting process for this model seemed inadequate in view of the MSE values which ran between 88 and 100, in spite of the consideration of depolarization in the fitting model. In figure \ref{depolarization} we show an example of a representative depolarization spectra considered in the optical fitting model.\\

When using SE and considering an anisotropic material two refractive indices per wavelength are obtained from the software, the ordinary $n_{ord-SE}$ and the extraordinary $n_{ext-SE}$. The average of these values is an effective refractive index $n_{ef-SE}$ for any polarization. Here we consider both options for normal incidence, the $n_{ef-SE}$ and $n_{ord-SE}$ in order to test them. Thus using these values we fabricated, as described in the main text, the simplest 1D photonic structures: named Bragg reflectors. By means of the $Q_e$- factor we evaluated the refractive indices and obtained the following values for i) the mirrors fabricated with $n_{ef-SE}$ : $Q_{e600}= 0.7908$, $Q_{e800}= 0.6583$ and $Q_{e1000}= 0.6065$, where the subscript stands for the central wavelength  of each Bragg reflector respectively. ii) the mirrors fabricated with $n_{ord-SE}$ : $Q_{e600}= 0.6831$, $Q_{e800}= 0.5928$ and $Q_{e1000}= 0.5564$. The $Q_e$- factors obtained from the Bragg reflectors fabricated with the ordinary refractive indices are smaller than the ones obtained from the effective refractive indices where a contribution of the $n_{ext-SE}$ is taken into account. Based on these results we used the ordinary refractive indices to compare them to other possibilities. These are the values we report in the main text under SE.
\begin{figure}[ht]
\centering
\includegraphics[width=0.8\columnwidth]{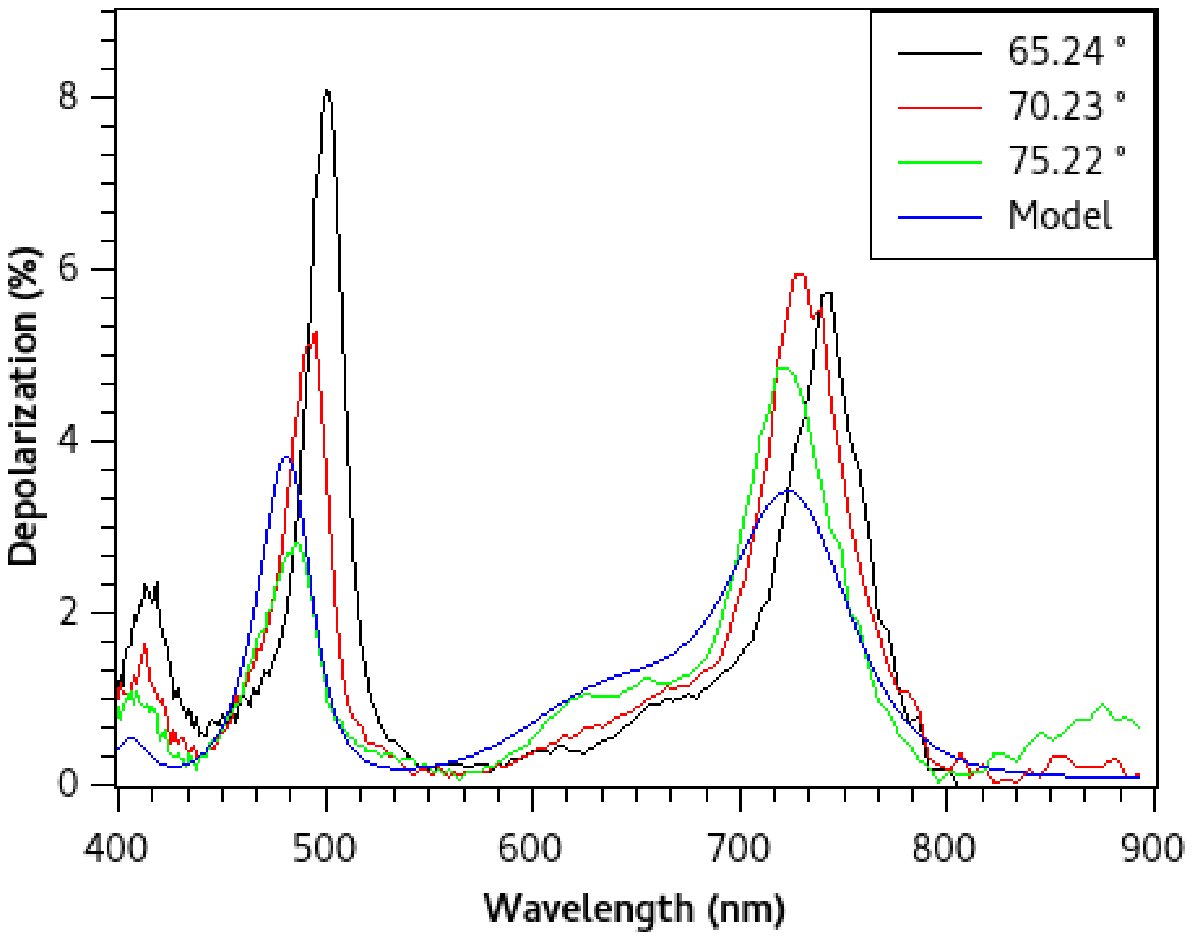}
\caption{Representative depolarization spectra of one of the high porosity PS layers. } 
\label{depolarization}
\end{figure}

\begin{figure}[H]
\centering
\includegraphics[width=1\columnwidth]{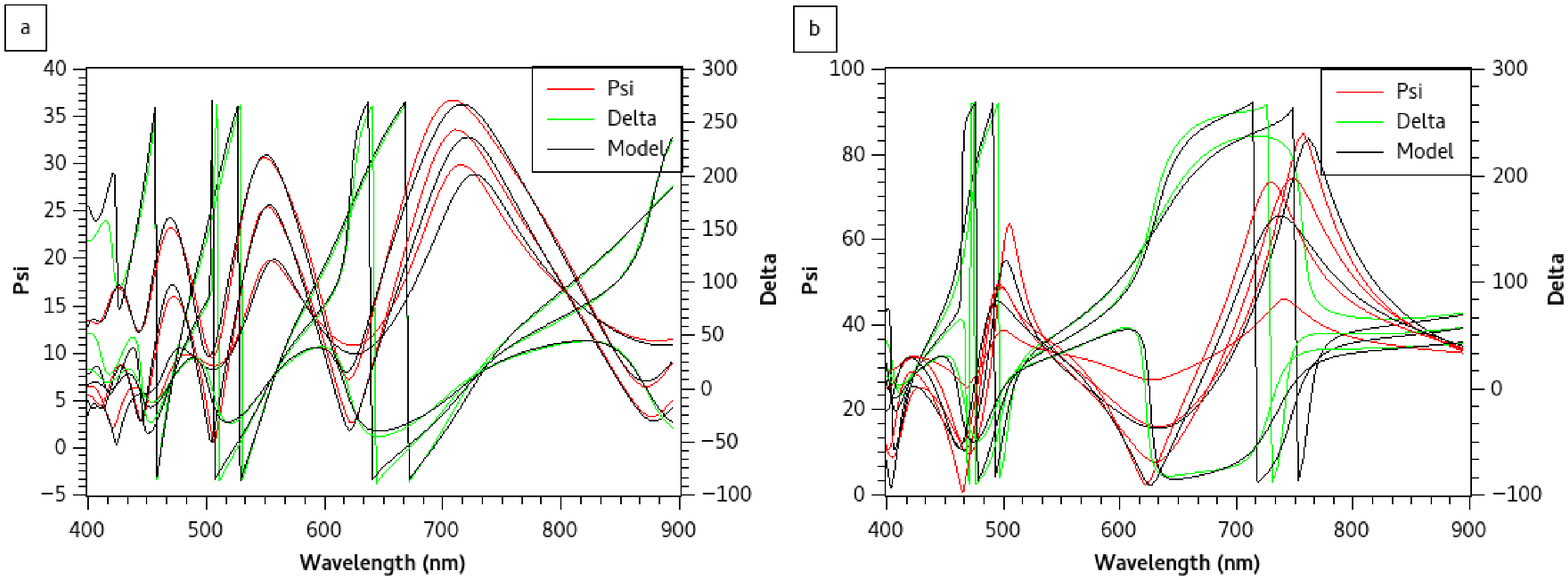}
\caption{ Ellipsometric measurements of fabricated PS layers of a) low porosity $P_b$ and b) high porosity $P_a$.} 
\label{angulos}
\end{figure}

In figure \ref{angulos} we show two representative ellipsometric angle measurements for the PS monolayers of high and low porosity using the optical models described above. The refractive indices of the PS films were obtained from the corresponding spectra and calculated their average values. The resulting refractive indices for the high and low porosity samples are plotted in figure [2] of the present work.\\

We would like to stress that there have been many optical models used for SE applied to a variety of different PS structures. Amongst them the gradual porosity and the anisotropy models are two commonly accepted approaches, which we used in this work based on previous reports \cite{Zeuz}. More importantly, these optical models were shown to be succesfull for the determination of refractive indices of PS used to fabricate photonic structures by Xifr\'e-P\'erez et al. However, following these approaches we did not obtain the expected results, since there was no agreement between the fabricated photonic structures and their theoretical counterpart. From our point of view, major analysis is required in order to understand the reasons why we were not able to find good agreement in the simplest photonic structures prepared with the SE refractive indices. 
In this work our main interest is to find adequate refractive indices to produce photonic structures, and we present reliable results when using a simple effective approximation method.

\section*{Appendix: Spectroscopic Ellipsometry measurements}
\setcounter{figure}{0}
\renewcommand{\thefigure}{A\arabic{figure}}

In this appendix we detail the procedure we used for the determination of the refractive index of high and low porosity PS samples by Spectroscopic Ellipsometry (SE). 

SE is a semiempirical approach based on the measurement of the light polarization transformation that occurs after a polarized beam is reflected from a material. The SE data are adjusted to fit a model that considers the characteristic properties of the material and from which the complex refractive index is calculated \cite{Zeuz,elipsometria}.

In this work we fabricated $500$ nm thick PS monolayers for the SE measurements, which were carried out on a $\alpha$-SE, J.A. Woollam Co spectroscopic ellipsometer. The data analysis was performed using the Complete EASE software where in the fitting model we established the PS as an effective Bruggeman medium (EMA) formed of Si and air. The parameters, such as thickness and porosity, are varied/fitted in the manner that the adjustment between simulated and experimental data present the minimum mean square error (MSE).

As it can be observed in the SEM image (figure 1 of the main text), the morphology of PS is complicated and varies with porosity. The structure of the low porosity layers resembles interlaced branches of Si, whereas the high porosity layers have a coral-like formation which present percolation. These different structures might be represented by different fitting models, such as isotropic, anisotropic or porosity graded. In this work we tried the above mentioned possibilities for the $P_b$ layers and found the minor MSE values using the graded porosity model. Because of the coral-like structure observed in the $P_a$ layers we used an anisotropic model for the fitting of these layers. The optical graded and anisotropic models are commonly used to approach the optical properties of PS when using SE \cite{Zeuz, elipsometria} and in this study we used them as follows:

a) For the low porosity layers we defined a depth dependent porosity gradient by simulating $10$ isotropic sublayers with a gradual porosity. This graded layer is on top of another layer with lower porosity and placed on the Si substrate. The fitting through the software gave a thickness of 493 nm on average, whereas from the SEM measurements we obtained a thickness of 511 nm. The MSE values obtained for this model range between 35 and 40.

b) For higher porosities we approached a coral-like structure where a directional dependency of the refractive index was generated and therefore considered an uniaxial anisotropic layer \cite{Zeuz}. The software gave a fitted thickness of 464 nm on average, which compared to the SEM measurements where we obtained a thickness of 507 nm, the SE fitting presented higher differences. 
The fitting process for this model seemed inadequate in view of the MSE values which ran between 88 and 100, in spite of the consideration of depolarization in the fitting model. In figure \ref{depolarization} we show an example of a representative depolarization spectra considered in the optical fitting model.\\

When using SE and considering an anisotropic material two refractive indices per wavelength are obtained from the software, the ordinary $n_{ord-SE}$ and the extraordinary $n_{ext-SE}$. The average of these values is an effective refractive index $n_{ef-SE}$ for any polarization. Here we consider both options for normal incidence, the $n_{ef-SE}$ and $n_{ord-SE}$ in order to test them. Thus using these values we fabricated, as described in the main text, the simplest 1D photonic structures: named Bragg reflectors. By means of the $Q_e$- factor we evaluated the refractive indices and obtained the following values for i) the mirrors fabricated with $n_{ef-SE}$ : $Q_{e600}= 0.7908$, $Q_{e800}= 0.6583$ and $Q_{e1000}= 0.6065$, where the subscript stands for the central wavelength  of each Bragg reflector respectively. ii) the mirrors fabricated with $n_{ord-SE}$ : $Q_{e600}= 0.6831$, $Q_{e800}= 0.5928$ and $Q_{e1000}= 0.5564$. The $Q_e$- factors obtained from the Bragg reflectors fabricated with the ordinary refractive indices are smaller than the ones obtained from the effective refractive indices where a contribution of the $n_{ext-SE}$ is taken into account. Based on these results we used the ordinary refractive indices to compare them to other possibilities. These are the values we report in the main text under SE.
\begin{figure}[ht]
\centering
\includegraphics[width=0.8\columnwidth]{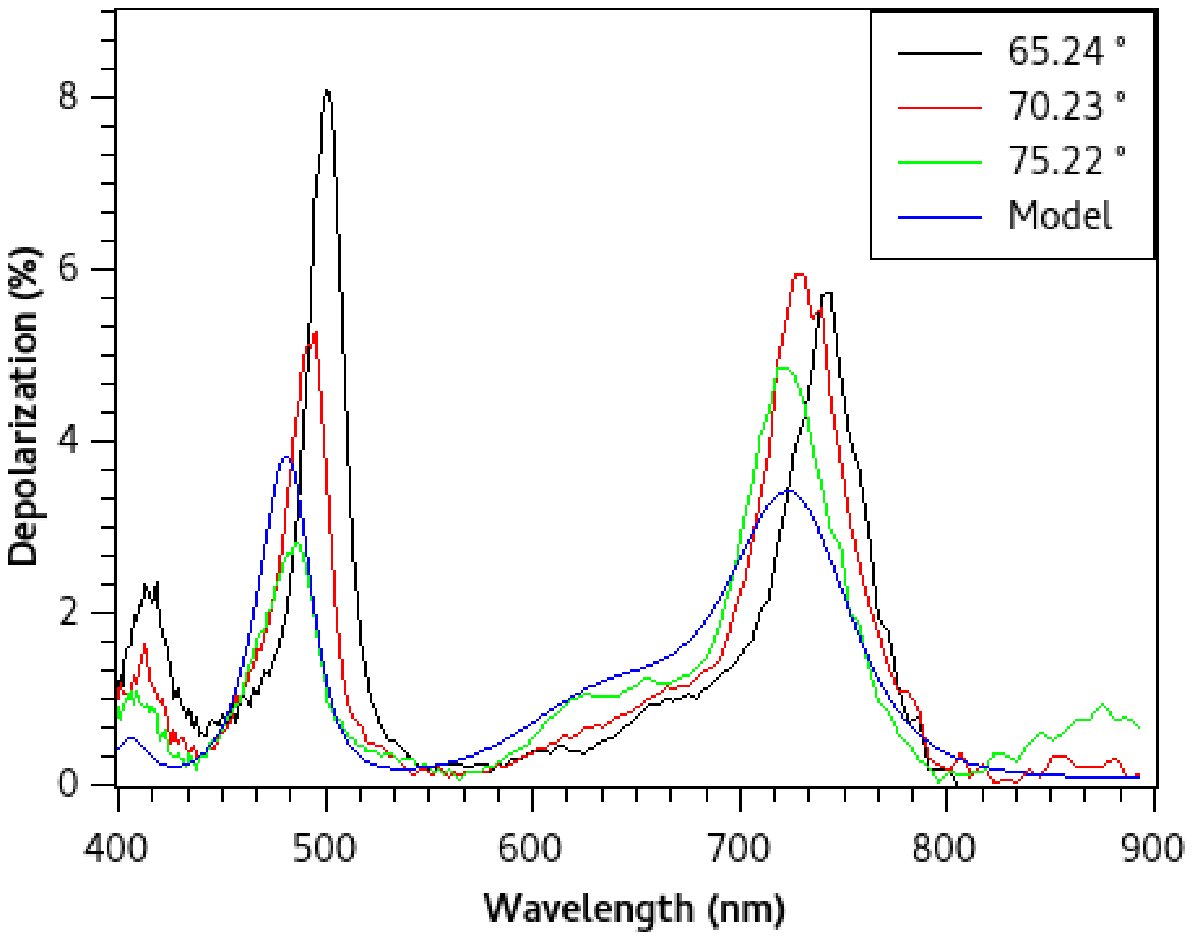}
\caption{Representative depolarization spectra of one of the high porosity PS layers. } 
\label{depolarization}
\end{figure}

\begin{figure}[H]
\centering
\includegraphics[width=1\columnwidth]{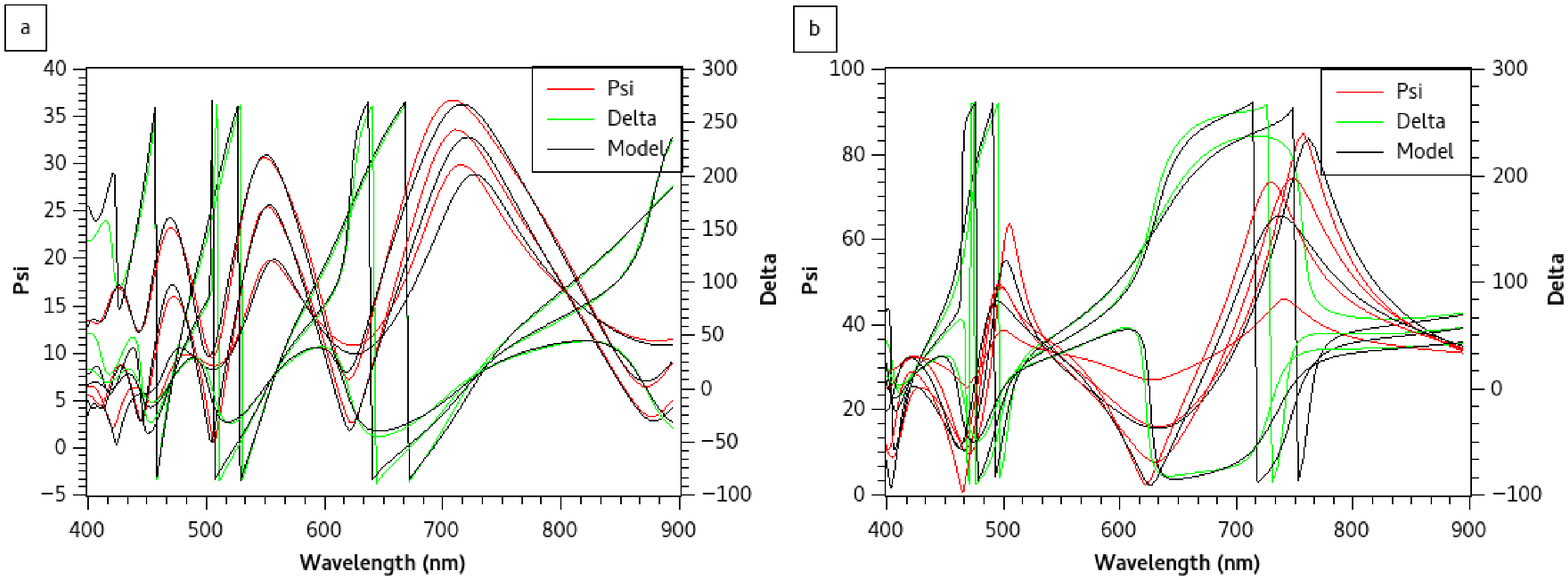}
\caption{ Ellipsometric measurements of fabricated PS layers of a) low porosity $P_b$ and b) high porosity $P_a$.} 
\label{angulos}
\end{figure}

In figure \ref{angulos} we show two representative ellipsometric angle measurements for the PS monolayers of high and low porosity using the optical models described above. The refractive indices of the PS films were obtained from the corresponding spectra and calculated their average values. The resulting refractive indices for the high and low porosity samples are plotted in figure [2] of the present work.\\

We would like to stress that there have been many optical models used for SE applied to a variety of different PS structures. Amongst them the gradual porosity and the anisotropy models are two commonly accepted approaches, which we used in this work based on previous reports \cite{Zeuz}. More importantly, these optical models were shown to be succesfull for the determination of refractive indices of PS used to fabricate photonic structures by Xifr\'e-P\'erez et al. However, following these approaches we did not obtain the expected results, since there was no agreement between the fabricated photonic structures and their theoretical counterpart. From our point of view, major analysis is required in order to understand the reasons why we were not able to find good agreement in the simplest photonic structures prepared with the SE refractive indices. 
In this work our main interest is to find adequate refractive indices to produce photonic structures, and we present reliable results when using a simple effective approximation method.

\end{document}